\documentclass[aps,twocolumn, prb, groupedaddress, showpacs, showkeys]{revtex4}
\usepackage{graphicx}
\usepackage{amsmath}

\begin{document}


\title{Epitaxial growth of Fe$_{3}$O$_{4}$ thin films on ZnO and MgO substrates}


\author{A. M\"uller}
\email[]{andreas.mueller@physik.uni-wuerzburg.de}
\author{A. Ruff}
\author{M. Paul}
\author{A. Wetscherek}
\author{G. Berner}
\affiliation{Universit\"at W\"urzburg, Experimentelle Physik 4, D-97074 W\"urzburg, Germany}
\author{C. Praetorius}
\author{K. Fauth}
\affiliation{Universit\"at W\"urzburg, Physikalisches Institut, D-97074 W\"urzburg, Germany}
\author{U. Bauer}
\author{M. Przybylski}
\affiliation{MPI f{\"u}r Mikrostrukturphysik, Weinberg 2, D-06120 Halle, Germany}
\author{M. Gorgoi}
\affiliation{Helmholtz Zentrum Berlin (BESSY II), Albert-Einstein-Str. 15, D-12489 Berlin, Germany}
\author{M. Sing}
\author{R. Claessen}
\affiliation{Universit\"at W\"urzburg, Experimentelle Physik 4, D-97074 W\"urzburg, Germany}


\date{\today}

\begin{abstract}
Magnetite (Fe$_{3}$O$_{4}$) thin fims have been grown epitaxially on ZnO and MgO substrates using molecular beam epitaxy. The film quality was found to be strongly dependent on the oxygen partial pressure during growth. Structural, electronic, and magnetic properties were analyzed utilizing Low Energy Electron Diffraction (LEED), HArd X-ray PhotoElectron Spectroscopy (HAXPES), Magneto Optical Kerr Effect (MOKE), and X-ray Magnetic Circular Dichroism (XMCD). Diffraction patterns show clear indication for growth in the (111) direction on ZnO. Vertical structure analysis by HAXPES depth profiling revealed uniform magnetite thin films on both type of substrates. Both, MOKE and XMCD measurements show in-plane easy magnetization with a reduced magnetic moment in case of the films on ZnO.

\end{abstract}

\pacs{79.60.Dp, 79.60.Jv}
\keywords{magnetite, zinc oxide, Fe$_{3}$O$_{4}$, ZnO, photoemission, leed, xmcd, haxpes}

\maketitle

\section{Introduction}
The realization of spintronic devices, such as spin-transistors or spin-valves, strongly depend on the availability of new materials that combine both ferromagnetic and semiconducting properties.\cite{awschalom2007} Besides the incorporation of magnetic impurities into a semiconducting host as in the so-called diluted magnetic semiconductors, the controlled injection of a spin-polarized current using ferromagnet-semiconductor heterostructures is a promising approach.\cite{ohno1998, wolf2001} Applications in this field are, e.g., spin-polarized light-emitting diodes and lasers.\cite{holub2007, fiederling1999} Fe$_{3}$O$_{4}$/ZnO hybrid systems should suit this purpose especially well. Semi-metallic magnetite (Fe$_{3}$O$_{4}$) is a ferrimagnet and predicted to possess a fully minority spin polarized Fermi surface. Since its N\'{e}el temperature lies well above ambient temperature it is one of the most promising candidates as a spin-injector.\cite{haghiri2004} Zinc oxide (ZnO) attracts attention for its application in transparent opto-electronics since it exhibits a wide direct band gap.\cite{look2001}

In this paper we report on the first epitaxial growth of Fe$_{3}$O$_{4}$ thin films on zinc oxide (ZnO) by molecular beam epitaxy (MBE). As reference system we also show data on the Fe$_{3}$O$_{4}$/MgO system which is well documented in the literature.\cite{ruby1999, voogt1999, lazarov2005, handke2001}

\section{Material requirements and properties}

\begin{figure*}
    \begin{minipage}{0.45\textwidth}
    \includegraphics[width=1\textwidth]{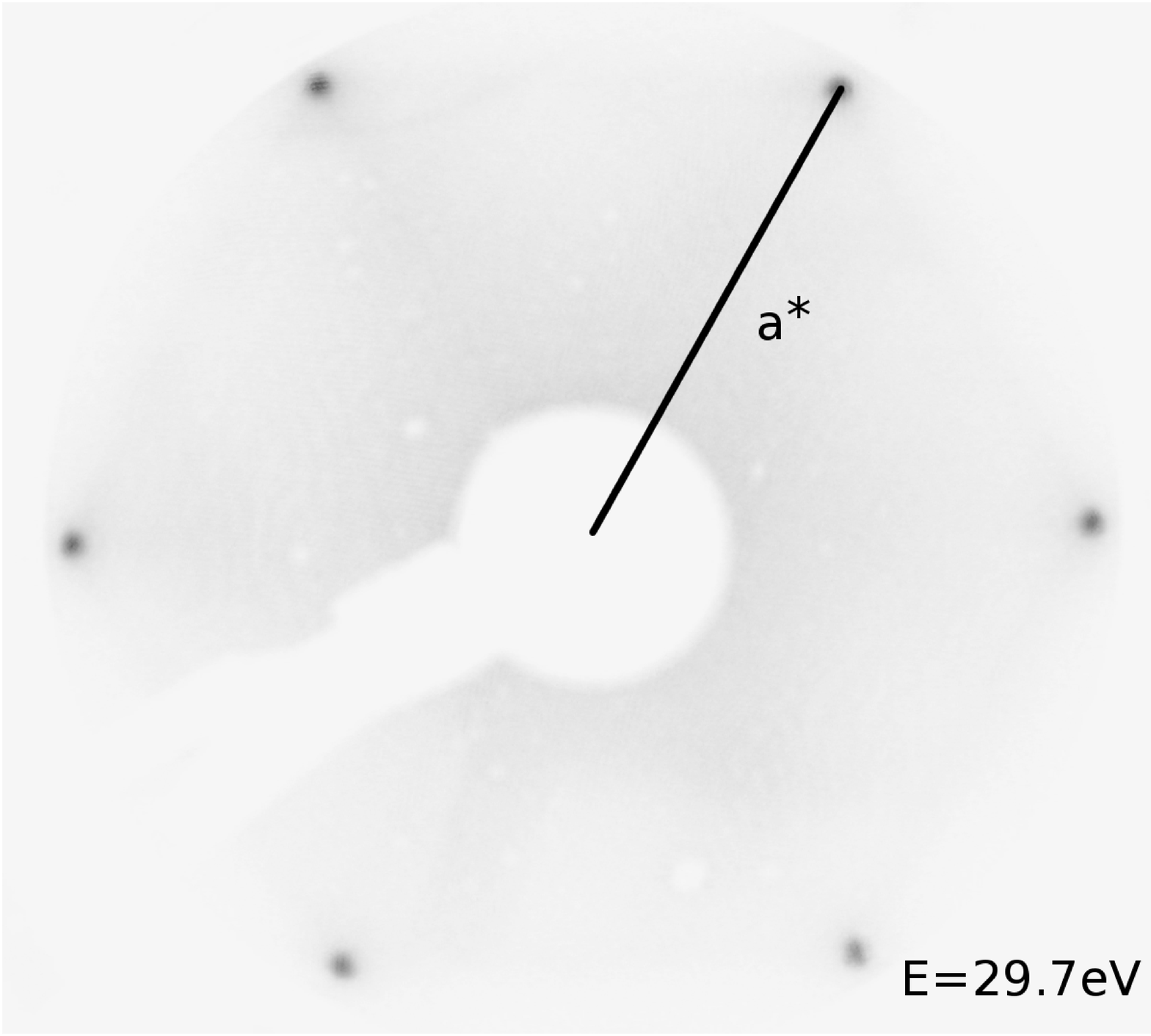}
    \caption{\label{fig:LEEDZnO}LEED pattern (E=29.7\,eV) of a ZnO substrate showing a six-fold symmetry. The lattice constant is calculated to be 3.3$\pm$0.2\,\AA. The low background intensity indicates a clean, well-ordered substrate surface.}
    \end{minipage}
    \hspace{0.3cm}
    \begin{minipage}{0.45\textwidth}
    \includegraphics[width=1\textwidth]{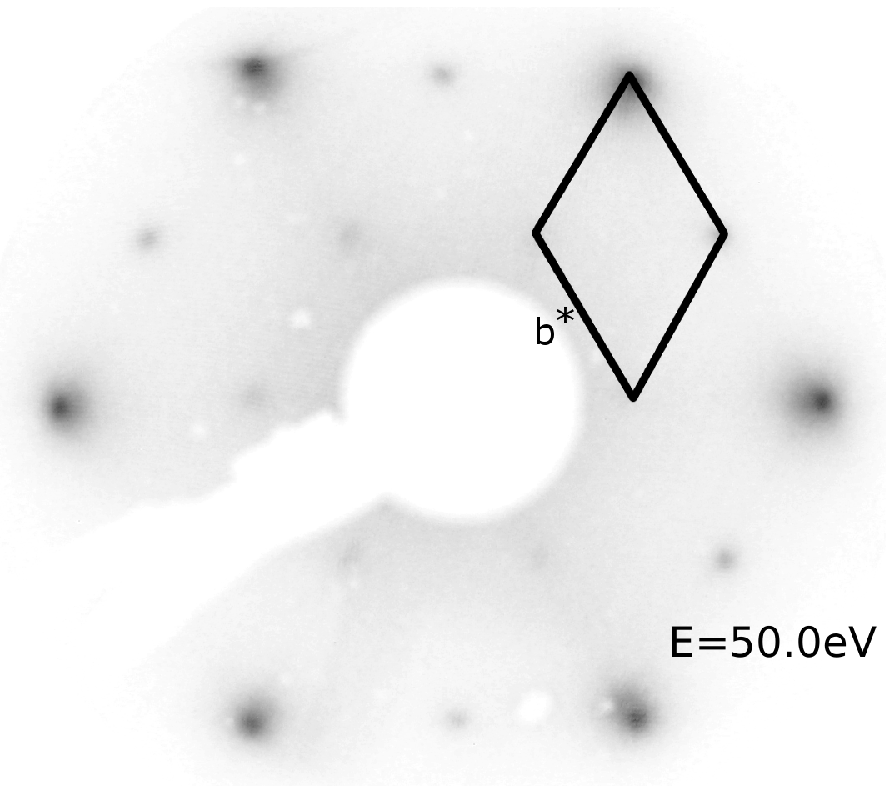}
    \caption{\label{fig:LEEDFe3O4}LEED pattern (E=50.0\,eV) of a Fe$_{3}$O$_{4}$ thin film grown on ZnO substrate. The hexagonal symmetry implies a growth in (111) direction of the Fe$_{3}$O$_{4}$ film. The surface exhibits no reconstruction, but reasonably sharp spots and low background intensity hinting to single crystal quality.}
    \end{minipage}
\end{figure*}

There are some basic requirements for ferromagnetic-semiconducting heterostructures that need to be met. (i) In addition to a maximum spin-polarization in the ferromagnet, the impedance mismatch of both materials must be as small as possible to accomplish efficient spin-injection.\cite{schmidt2000} (ii) Spin scattering lengths in both materials will sensitively depend on intrinsic properties of each compound, device geometry, and the crystalline quality of the heterostructure.\cite{holub2007} I.e.,~homogeneous film formation strongly depends on growth parameters like temperature, background pressure, and lattice mismatch. This latter point we will mainly discuss in our report.

As spin-aligning material, the iron oxide Fe$_{3}$O$_{4}$ stands out with a theoretically -100\,\% spin-polarization at the Fermi energy and a very high N\'{e}el temperature of T$_{N}$=858\,K.\cite{yanase1999,samara1969} However, depending on the crystal quality and surface orientation experiments show a spin-polarization only between -55\,\% and -80\,\%.\cite{fonin2005, dedkov2002} The ferrimagnetic behavior results from anti-ferromagnetic coupling of the spins in the two Fe sublattices of the inverse spinel Fe$_{3}$O$_{4}$ crystal structure. Theoretical and experimental analysis derive a net magnetization of 4.0$\mu_{B}$ per formula unit.\cite{kakol1989, zhang1991} The electronic conductivity of 2.5$\cdot$10$^{-4}$\,($\Omega$cm)$^{-1}$ at room temperature is reasonably well within the limit for spin-injection into a semiconductor material.\cite{miles1957}

ZnO is currently explored due to its semiconducting properties (E$_{gap}$=3.37\,eV) and its potential application in various oxide-electronic devices.\cite{pearton2003, look2001}
Despite the large lattice mismatch (a$_{ZnO}$=3.25\,\AA) with respect to the Fe$_{3}$O$_{4}$ (111) surface (b$_{111}$=5.92\,\AA), the oxidic character itself and first reports of the successful growth of Fe$_{3}$O$_{4}$ films on ZnO by pulsed laser deposition justify the attempt to grow these films by MBE.\cite{pearton2003, condon1996, nielson2008}

\section{Growth and \textit{in situ} characterization by LEED}

\begin{figure*}
    \begin{minipage}{0.45\textwidth}
    \includegraphics[width=1\textwidth]{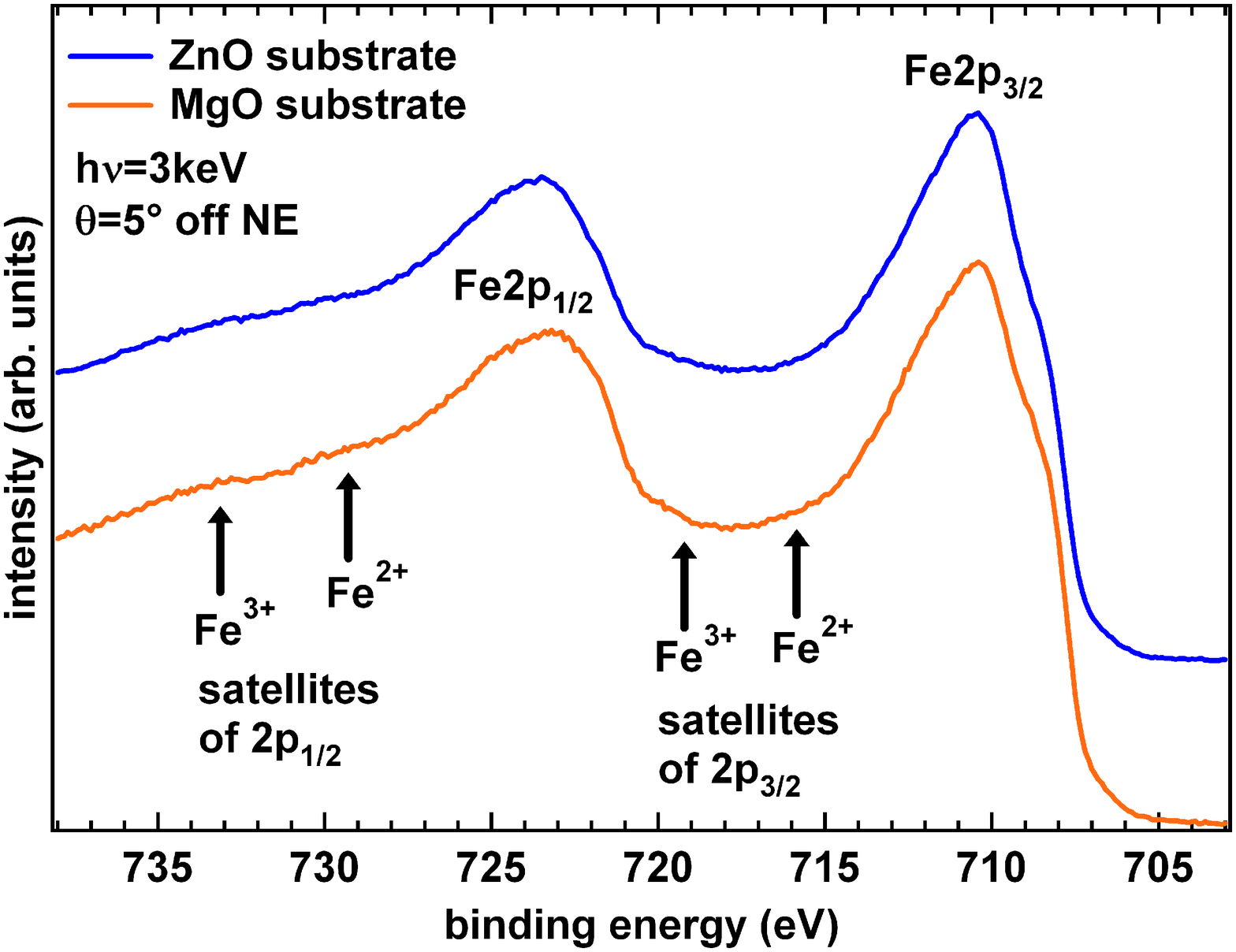}
    \caption{\label{fig:CompFe2p}(Color online) HAXPES spectra of the Fe~$2p$ core-level. Upper line (blue): film grown on ZnO; lower line (orange): film grown on MgO.  The positions of the charge transfer satellites for the Fe$^{2+}$ and Fe$^{3+}$ valence states are indicated. Both spectra show the characteristic shape of a mixed Fe$^{2+}$/Fe$^{3+}$ valence state as is characteristic for Fe$_{3}$O$_{4}$.}
    \end{minipage}
    \hspace{0.3cm}
    \begin{minipage}{0.45\textwidth}
    \includegraphics[width=1\textwidth]{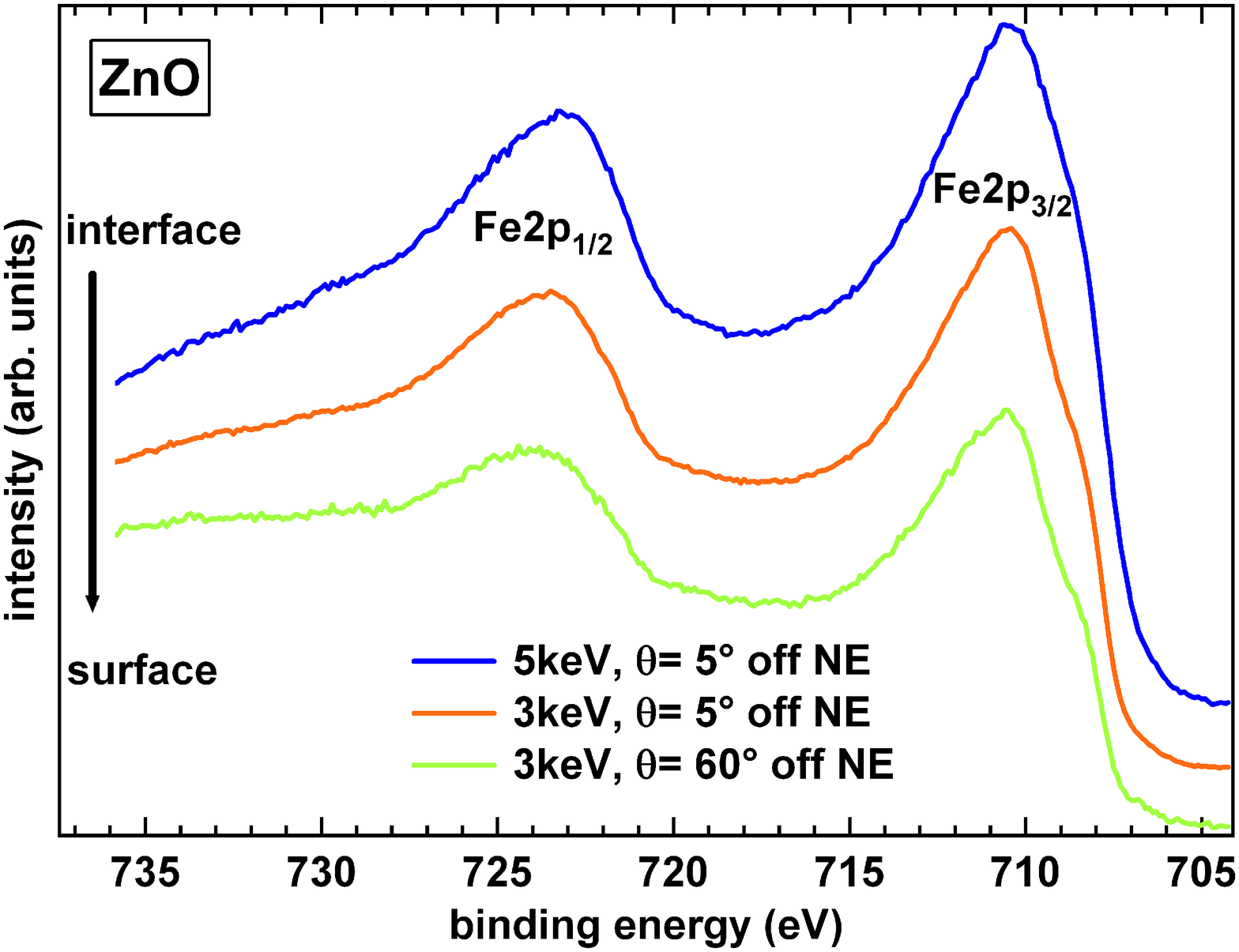}
    \caption{\label{fig:DepthFe2p}(Color online)  Vertical profiling of the chemical structure using HAXPES: Fe~$2p$ spectra of a Fe$_{3}$O$_{4}$ thin film on ZnO substrate. The spectra show no dependence on photon energy and detection angle indicating a homogenous film. Neither the interface sensitive spectra (blue, upper line) nor the surface sensitive spectra (green, lower line) show any other Fe oxidation state.}
    \end{minipage}
\end{figure*}

There are two different approaches to grow iron oxide films epitaxially by MBE. First, pure iron thin films can be grown on substrates under ultra high vacuum conditions followed by post-oxidation at the appropriate temperatures and oxygen partial pressures. Another approach is the growth \textit{in} oxygen atmosphere. In this case the desired oxidation is already achieved during growth and a more uniform film growth could be expected. Furthermore, oxygen vacancies in the substrate caused by reduction during heating in vacuum or due to oxygen diffusion from substrate to film is avoided. If substrate oxidation does not play a role, or even oxidic substrates are used, the second approach is the method of choice which we used for both Fe$_{3}$O$_{4}$/ZnO and Fe$_{3}$O$_{4}$/MgO hybrid systems.

For film growth an electron beam evaporator equipped with an ultra-high purity iron rod was used. The oxygen pressure was controlled by a leak valve, applying molecular oxygen. Growth temperature was monitored \textit{ex situ} with a pyrometer. \textit{In situ} characterization was done using LEED and x-ray photoelectron spectroscopy (XPS).

Films were grown on ZnO(0001) and MgO(111) (Mateck, Germany) substrates. Both substrates were cleaned \textit{ex situ} using organic solvents. \textit{In situ} treatment included several cycles of ion-etching and annealing up to T=975\,K. As last step, both substrates were annealed at T=775\,K and in an oxygen partial pressure of p(O$_{2}$)=5$\cdot$10$^{-7}$\,mbar for 15\,min.

The Fe$_{3}$O$_{4}$ film on MgO was grown at T=580\,K and an oxygen partial pressure of p(O$_{2}$)=2.5$\cdot$10$^{-7}$\,mbar. From measurements of the iron flux via a calibrated fluxmeter in the electron beam evaporator film thickness was estimated to be 18$\pm$4\,nm. The magnetite film on ZnO was grown at the same temperature. The oxygen pressure was kept at p(O$_{2}$)=5$\cdot$10$^{-6}$\,mbar for the first 10\,min and reduced to p(O$_{2}$)=1$\cdot$10$^{-6}$\,mbar for the remaining growth duration (30\,min) to prevent over-oxidation (see Fig.\,\ref{fig:layer}). Iron flux measurements gave a thickness of about 12$\pm$3\,nm. Both samples were cooled down to room temperature without any oxygen dosing.

A typical LEED pattern of an \textit{in situ} treated ZnO substrate is shown in Fig.\,\ref{fig:LEEDZnO}. The image is taken at an electron energy of E=29.7\,eV. Similar patterns could be obtained over a wide energy range. The pattern exhibits six-fold symmetry with clear main spots and low background intensity. This, together with the absence of additional spots indicate a clean unreconstructed surface with a low defect rate and small surface roughness. Analyzing this pattern, a lattice constant of a=3.3$\pm$0.2\,\AA was calculated, which fits perfectly well to literature values for the unreconstructed ZnO(0001) surface lattice constant.\cite{pearton2003} The substrate was also checked by XPS (spectra not shown) which showed no contaminations after treatment.

Figure\,\ref{fig:LEEDFe3O4} shows the LEED pattern of a Fe$_{3}$O$_{4}$ film on ZnO, taken at an energy of E=50.0\,eV. Like the substrate pattern it exhibits a six-fold symmetry. Clearly visible is a variation in spot intensity and width which was not present in the substrate pattern. This pattern of differently bright spots is commonly ascribed to the unreconstructed Fe$_{3}$O$_{4}$(111) surface.\cite{berdunov2004, ritter1999} The low background intensity and comparably sharp spots, that could be obtained all over the sample, indicate epitaxial growth and a long range order of the film. However, in comparison to LEED patterns of Fe$_{3}$O$_{4}$ on MgO that are well documented in the literature, spots are less sharp, and no reconstructions were found for the films on ZnO.\cite{voogt1999, handke2001} This could be a hint to a slightly increased surface roughness for the films on ZnO, which indeed could be confirmed by AFM images (not shown). The hexagonal symmetry clearly suggests a growth in (111) direction. As a guide for the eye the unit cell is sketched in the pattern. The lattice constant \textit{b} can be determined to be 5.96$\pm$0.2\,\AA~which compares well with values for the (111) surface in the literature.\cite{condon1996, paul2008}

\section{Electronic and chemical depth profiling}

HAXPES measurements were performed at the \mbox{KMC-1} beamline at Helmholtz Zentrum Berlin (BESSY II) using the HIKE endstation. All samples were transferred under argon atmosphere between the vacuum systems, with exposure to ambient atmosphere being kept as short as possible. Measurements were performed without any further surface treatment at room temperature.

HAXPES measurements of the core-levels allow for a detailed study of the chemical composition. In addition, it is sensitive to the oxidation/valence state. Using photons in the hard X-ray regime substrate, interface, and film can be probed as the electron escape depth $\lambda$ scales with the kinetic energy of the photoelectrons. The effective electron escape depth also depends on the electron detection angle $\theta$. Thus non-destructive depth profiling is possible, if either of these two parameters (detection angle and photon energy) is varied. Film thicknesses and sample compositions can be determined by a quantitative analysis of the spectra. For depth profiling, spectra were measured using photon energies of 3, 4, and 5\,keV and electron detection angles of $\theta$=5$^\circ$, $\theta$=20$^\circ$, 40$^\circ$, and 60$^\circ$ off normal emission. This results, e.g. for the Fe~$2p$ core-level spectra, in an information depth \textit{d=3$\cdot\lambda(E,\theta)$} between 5.7\,nm and 18.9\,nm using the formula given by Tanuma \textit{et al.}~\cite{tanuma2005, nist} Comparing this values to the film thicknesses, we can extract the properties of the interface, film, and surface from the measured spectra.

An energy resolution of 0.51\,eV at a photon energy of 3\,keV was determined by measuring the Au\,4\textit{f$_{7/2}$} core-levels. All spectra were calibrated to the O~$1s$ core-level at 530.1\,eV, as the binding energy for this particular core-level is the same in all iron oxides and well documented in the literature.\cite{gota1999}

\begin{figure}
    \includegraphics[width=0.45\textwidth]{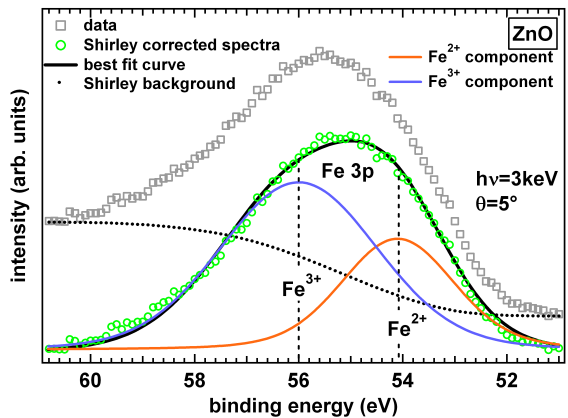}
    \caption{\label{fig:Fe3p}(Color online)  Quantitative analysis of the Fe~$3p$ core-level spectrum for films grown on ZnO. A least squares fit using two Voigt profiles gives an Fe$^{3+}$\,:\,Fe$^{2+}$ area ratio of $(1.9\,\pm0.2):~1$.}
\end{figure}

HAXPES spectra of the Fe~$2p$ core-level are displayed in Fig.\,\ref{fig:CompFe2p} and Fig.\,\ref{fig:DepthFe2p}. The spectra show the spin-orbit split Fe~$2p_{3/2}$ and Fe~$2p_{1/2}$ component at lower (710.5\,eV) and higher binding energies (723.6\,eV), respectively. Remarkable is the shoulder at the lower binding energy side of the spectrum (708.8\,eV and 721.9\,eV, respectively).
Note the flat region between the spin-orbit split components.

In Fig.\,\ref{fig:CompFe2p} the Fe~$2p$ spectra of films grown on MgO (orange, lower line) and ZnO (blue, upper line) are shown. Indicated are the positions of the Fe~2$p$ charge transfer satellites
that depend on the Fe valence. For example, for the Fe~$2p_{3/2}$ peak the Fe$^{2+}$ satellite occurs at 715.6\,eV, whereas the Fe$^{3+}$ appears at 719.2\,eV. In purely di- or trivalent Fe compounds they are clearly visible, however, in mixed-valent Fe$_3$O$_4$ they add up to the nearly flat region between the 2p$_{3/2}$ and 2p$_{1/2}$ main lines, thus providing a clear spectral signature for magnetite.\cite{gota2001,ruby1999,yamashita2008}

The chemical shift that is induced by the different Fe valences (Fe$^{2+}$/Fe$^{3+}$, $\Delta$E=1.7\,eV), results in a shoulder at the lower BE side of the main lines as already mentioned above. Due to its absolute position and its relative shift with respect to the main line, this shoulder can be clearly attributed to Fe$^{2+}$, whereas the main peak corresponds to Fe$^{3+}$.\cite{gota2001} A possible non-oxidized pure iron component can be excluded, as this would result in an additional feature around E=707.0\,eV for the Fe~$2p_{3/2}$ peak, which is not present in our spectra.\cite{gota2001, wett2008} Both films, on MgO and ZnO, exhibit the same features and are virtually identical.

\begin{figure*}
    \includegraphics[width=0.99\textwidth]{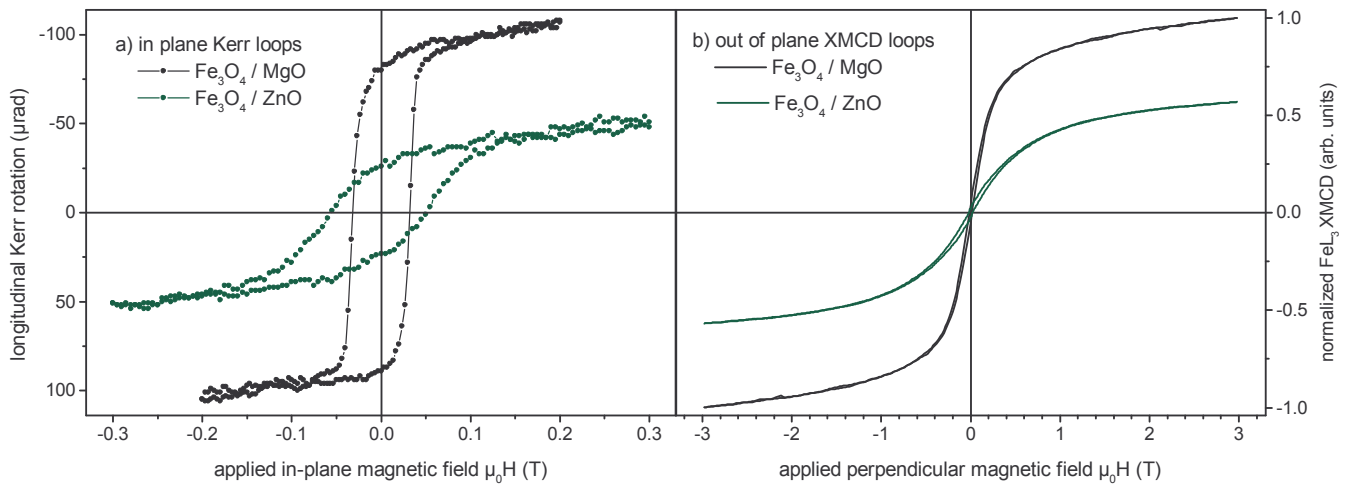}
    \caption{\label{fig:magncurv}(Color online)  Magneto-optically detected magnetization curves of Fe$_3$O$_4$ films grown on MgO and ZnO, respectively. a) l-Kerr loops sensitive
    to in-plane magnetization. Note the negative Kerr rotation. b) Normal incidence XMCD loops sensitive to perpendicular magnetization. Ordinate rescaled to the sum rule results obtained at $\pm 3$ T (see text).}
\end{figure*}

In Fig.\,\ref{fig:DepthFe2p}, the Fe~$2p$ spectra of an iron oxide film on ZnO is shown for different measuring geometries and photon energies. Thereby, the surface sensitivity increases from the upper to the lower spectrum. The blue line (upper spectrum) represents the most interface sensitive spectrum. In all spectra both core-levels are at the same binding energy and also the peak shape is identical. Comparing the three measurement geometries it is quite obvious that also in the vertical direction of the film, going from the interface to the surface, iron is in the same oxidation state. A higher background intensity on the higher binding energy side is observable for more surface sensitive measuring geometries, which is due to different detection angles and photon energies used.

Figure\,\ref{fig:Fe3p} shows the Fe~$3p$ core-level spectrum of a film grown on ZnO. The spectrum (squares) was measured using a photon energy of 3\,keV and a detection angle of $\theta$=5$^\circ$. After Shirley background correction (Shirley background: dotted line, Shirley corrected data: circles) a least squares fit procedure was applied using two Voigt profiles corresponding to Fe$^{2+}$ and Fe$^{3+}$.\cite{shirley1972} The resulting peaks are at a binding energy of 56.0\,eV and 54.1\,eV. The charge transfer satellites at the higher BE side have negligible intensity for this core-level and were not included in the fitting procedure. The relative shift of the two fitted Voigt profiles of 1.9\,eV is quite consistent with experimental results from McIntyre \textit{et al.}\cite{mcintyre1977} The fitting procedure yields a Fe$^{3+}$\,:\,Fe$^{2+}$ ratio of $(1.9\,\pm0.2):\,1$, which agrees well with the value expected for Fe$_{3}$O$_{4}$. The error can be estimated to be very small, since no assumptions for cross-sections or inelastic mean free path of the photoelectrons are necessary.

We were also able to determine the Fe$_3$O$_4$ film thickness by the damping of the Zn~$2p{_3/2}$ signal (not shown), using the standard formula $I=I_{0}\cdot\exp(-d/\lambda(E)\cos \theta)$.\cite{seah1990} Here $\lambda(E)$ is the inelastic mean free path of the photoelectrons, $\cos \theta$ the detection angle of the electrons, $d$ the film thickness, $I$ and $I_{0}$ the intensity of the Zn~$2p_{3/2}$ core-level with and without iron oxide film. Thickness was determined to be 10.9$\pm$1\,nm confirming the value derived by flux meter measurements.

\section{Magnetic properties}
Magnetic properties of the films were investigated near 300\,K by magneto-optical Kerr effect (MOKE) and x-ray magnetic circular dichroism (XMCD), respectively. MOKE experiments were conducted at a photon energy of $\hbar \omega = 1.85$\,eV and applied fields of up to $\mu_0 H_{max}=\pm0.7$\,T in both polar (p-MOKE) and longitudinal (l-MOKE) geometry (at $69^{\circ}$ and $30^{\circ}$ incidence with respect to the surface normal, respectively) to probe in- and out-of-plane magnetization components. Measurements in longitudinal geometry (i.e.~magnetization loops) were recorded at various azimuthal orientations of the specimens in order to detect in-plane magnetic anisotropy. XMCD experiments were conducted at the PM 3 bending magnet beamline for circular polarization at Helmholtz Zentrum Berlin (BESSY II). Normal incidence x-ray absorption spectra (XAS) were recorded in the total electron yield mode (TEY, sample drain current) at constant helicity ($p_{circ} \approx 0.93$) and applied magnetic fields of $\mu_0 H = \pm 3$\,T. TEY magnetization curves were measured at the first (negative) extremum of the Fe L$_3$ XMCD spectrum.\cite{goering2000} To estimate the effect of TEY saturation additional XAS spectra were recorded with linear polarization and at various angles of incidence, ranging from normal incidence up to $70^\circ$ off normal.\cite{stoehr1992} From these datasets our estimate of the effective TEY escape depth amounts to $\lambda_{el}=1.8\pm0.5$\,nm which falls in the range of previously reported values (0.85 nm \cite{goering2006}, 5.0 nm \cite{gota2000}) and leads to moderate saturation corrections. At the same time, finite thickness effects in the TEY data can be safely excluded for our films.

Fig.~\ref{fig:magncurv}a displays in-plane magnetization l-MOKE loops of Fe$_{3}$O$_{4}$ films grown on both MgO and ZnO substrates. Despite their different texture the qualitative behavior as detected by MOKE is quite similar. Longitudinal Kerr ellipticity at $\hbar\omega=1.85$\,eV is negative in both types of specimens and from a comparison with p-MOKE loops (not shown) it follows that the surface normal corresponds to a hard magnetization axis. The shape of the l-MOKE loops proves nearly independent of azimuthal sample orientation. Thus, irrespective of the substrate our Fe$_{3}$O$_{4}$ films with $11-18$\,nm thickness display little, if any, in-plane magnetic anisotropy. At the quantitative level we observe a number of differences in the l-MOKE loops of the films grown on MgO and ZnO, respectively. ZnO films possess larger coercive fields ($\mu_0 H_C^{ZnO}\approx50$\,mT, $\mu_0 H_C^{MgO}\approx31$\,mT) and, at the same time, a reduced squareness of the hysteresis cycles. The absolute magnitude of the Kerr rotation is smaller by almost a factor of two with the ZnO based film which cannot be accounted for by a difference in film thickness alone. We rather conclude that magnetization is reduced in the ZnO based films and shall confirm this reasoning based on our XMCD measurements below.

\begin{figure*}
    \includegraphics[width=0.99\textwidth]{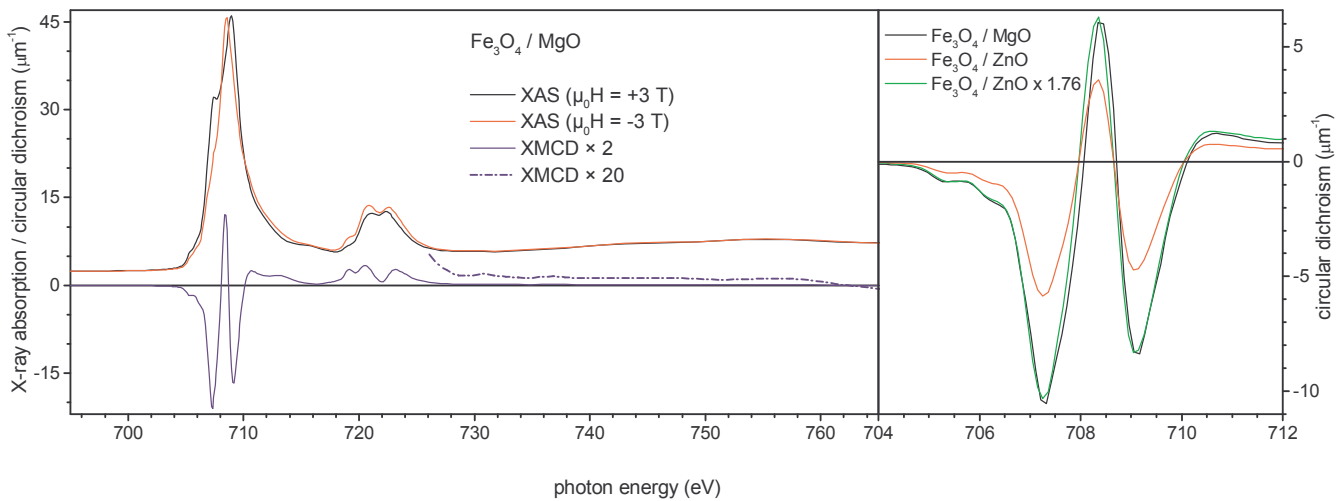}
    \caption{\label{fig:XMCD}(Color online)  Fe L$_{3,2}$ XAS and XMCD data of Fe$_3$O$_4$ grown on MgO and ZnO, respectively ($\mu_0H=\pm 3$\,T). a) Spectroscopic data for Fe$_3$O$_4$/MgO. A long integration range is required to reach convergence of the integrated dichroism.~\cite{goering2006} b) Comparison of Fe L$_3$ XMCD after growth on MgO and ZnO. While spectral shapes are similar, the XMCD response is substantially smaller on ZnO grown films (details see text).}
\end{figure*}

Fig.~\ref{fig:XMCD} displays the Fe L$_{3,2}$ absorption and circular dichroism data obtained from a film grown on MgO. The overall, polarization averaged absorption (not shown) compares favorably with available published data except for differences related to the various evaluations of $\lambda_{el}$.\cite{goering2006, regan2001} We cannot entirely exclude the formation of a slightly increased oxidation state, i.e.~the presence of some (minor) amount of $\gamma$-Fe$_{2}$O$_{3}$, in particular at the film surface. The dichroism spectrum, too, coincides well in its general features with those published by Goering \textit{et al.}~\cite{goering2006} In particular, we note that the circular dichroism signal extends more than 35\,eV above the L$_2$ absorption edge. From the comparison of our data with the XAS and XMCD spectra of Pellegrin \textit{et al.}~\cite{pellegrin1999}, we estimate the average stoichiometry of the films to be Fe$_{2.93 \pm 0.04}$O$_{4}$. Evaluation of the XMCD sum rules \cite{thole1992, carra1993} leads to average Fe $3d$ spin and orbital moments of $m_{S, eff} = 0.85 \mu_B$ and $m_L = 0.03 \mu_B$ per atom, respectively. In deriving these numbers, we have followed Goering \textit{et al.}~\cite{goering2006} in assuming an average Fe $3d$ electron count of 16.5 per formula unit and accounted for the finite circular polarization of the x-ray beam. From the TEY magnetization curves discussed above (see Fig.~\ref{fig:magncurv}b) it is evident that the saturation magnetization in our films is more elevated compared to Fe$_{3}$O$_{4}$ single crystals.

While the overall shapes of XAS spectra and spectral dichroism are very similar for Fe$_{3}$O$_{4}$ films grown on both MgO and ZnO substrates, the XMCD magnitude is considerably reduced in the latter case (see Fig.~\ref{fig:XMCD}b), which is in line with our previous finding of reduced Kerr rotation. Using this result, we can scale our XMCD-derived magnetization curves, measured at normal incidence and thus along the hard magnetization axes of the films (Fig.~\ref{fig:magncurv}b). As we expect for magnetization along the hard axis, both films possess reduced coercive fields in their out-of-plane loops compared to the in-plane loops. Nevertheless, as in our MOKE measurements, the ZnO based film displays a slightly larger coercive field (20\,mT vs.~15\,mT for the MgO based film). Both films are not magnetically saturated at the highest applied field of $\mu_0 H=3.0$\,T. The corresponding high field susceptibility, while larger for the film grown on MgO in absolute values, is nevertheless larger in case of the ZnO substrates when normalized by the magnetization. Thus, on a relative scale, Fe$_{3}$O$_{4}$ grown on MgO is more easily saturated than is Fe$_{3}$O$_{4}$ on ZnO. We shall further discuss this behavior below.

\section{Discussion}

\begin{figure}[b]
    \includegraphics[width=0.45\textwidth]{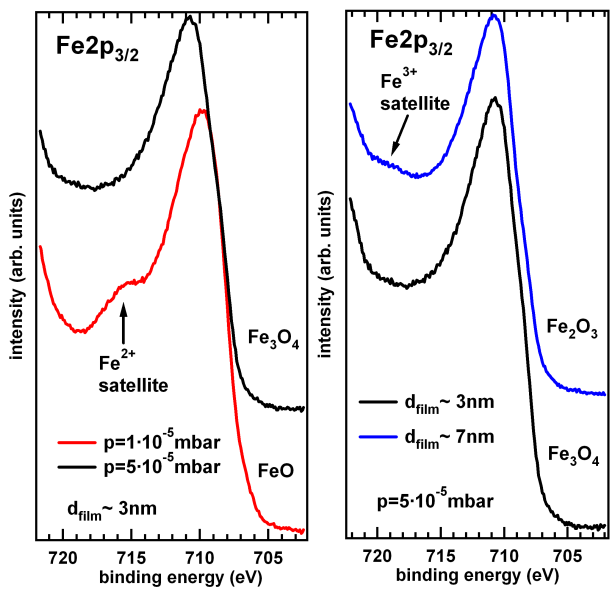}
    \caption{\label{fig:layer}(Color online) XPS spectra of the Fe~2p core-level of iron oxide films grown on ZnO. The spectra show that oxygen pressure and film thickness play an essential role in the formation of various Fe oxide phases.}
\end{figure}

The analysis of the HAXPES spectra and the LEED pattern of the film clearly suggest epitaxial growth of Fe$_{3}$O$_{4}$ on the ZnO substrate. Thereby, the in-plane lattice parameter seems to be fully relaxed towards the magnetite bulk value. The relaxation of the film seems to occur within the first 3\,nm as also LEED patterns of these thin films show this lattice constant. In terms of chemical composition and oxidation state, the magnetite film on ZnO shows the same level of quality compared to the film on MgO. Homogenous growth of the film in the vertical direction is confirmed by HAXPES depth profiling. The characteristic Fe~$2p$ line shape and the Fe$^{3+}$\,:\,Fe$^{2+}$ ratio of $(1.9\,\pm0.2):\,1$ are in line with the mixed Fe$^{3+}$\,:\,Fe$^{2+}$ valence state of Fe$_{3}$O$_{4}$.
We emphasize, that we could achieve such a uniform phase formation only by reducing the oxygen partial pressure during growth as described in section III. To clarify this point XPS spectra of differently prepared iron oxide films are shown in Fig.\,\ref{fig:layer}. The spectra in the left panel show the influence of the growth pressure on thin iron films. The oxygen pressure must be sufficiently high to force the formation of Fe$_{3}$O$_{4}$. A too low pressure results in the formation of FeO, which is identified by the Fe$^{2+}$ satellite peak in the XPS spectrum (red/lower line). It seems that at low pressure the formation of FeO is favorable due to the smaller lattice mismatch. For thicker films, however, the same high oxygen pressure leads to the formation of Fe$_{2}$O$_{3}$. This can be seen by the comparison of the spectra of two iron films that differ in thickness. The thicker film (blue/upper line) in the right panel of Fig.\,\ref{fig:layer} shows the appearance of the Fe$^{3+}$ satellite peak structure which indicates Fe$_{2}$O$_{3}$. By variation of the oxygen partial pressure during growth, uniform film formation can be achieved.

In line with the HAXPES results both XAS and XMCD confirm the similar stoichiometric quality of the Fe$_{3}$O$_{4}$ films grown on both MgO and ZnO substrates. The magnitude of the magnetic response is considerably lower for the films grown on ZnO consistently found with both MOKE and XMCD measurements. Since both kinds of films are not magnetically saturated at the maximum applied magnetic fields available to this study, questions concerning magnitude and the apparent substrate dependence of saturation magnetization cannot be settled at this time. However the experimental data reveal, that the films grown on ZnO are substantially harder to magnetize. Clearly, the magnetization of the ZnO based film lies below the one of its MgO counterpart at all fields. The normalized high field susceptibility $\left . M(H_{max})^{-1} \frac{dM}{dH}\right |_{H_{max}}$ is larger in the case of ZnO substrate, indicating a stronger departure from saturation magnetization. Since at the same time in- and out-of-plane coercive fields are larger with ZnO substrates it is plausible to look for a mechanism of magnetic anisotropy as the cause of these observations. Quite obviously, though, bulk-like magnetocrystalline anisotropy (MCA) cannot account for our observations, since no signs of significant MCA were found in the MOKE experiments.

Defects of various kinds including e.\,g.~antiphase boundaries are known to be local sources of potentially strong magnetic anisotropy \cite{celotto2003}. Locally, anisotropy at defects may be associated with spin-orbit-coupling at Fe sites with reduced symmetry. Our XMCD sum rule results do indeed provide evidence for the presence of a significant amount of such Fe sites. The ratio of orbital to spin magnetic moment derived from our XMCD measurements, $\frac{m_L}{m_{S, eff}} \approx 0.035$, is more than a magnitude larger than those found by Goering \textit{et al.}~at cleaved single crystal surfaces.\cite{goering2006} It appears reasonable, therefore, to associate the relative importance of magnetic anisotropy with the structural quality of the magnetite films. As indicated above, both LEED and atomic force microscopy characterization indicate an increased roughness of the ZnO based films compared to those grown on MgO. Accordingly, we conclude that the films grown on ZnO, while of similar stoichiometric quality as those grown on MgO, possess a higher density of defects resulting in both, a larger coercive field and a slower approach to saturation magnetization.

\section{Summary}
We successfully achieved the epitaxial growth of Fe$_{3}$O$_{4}$ thin films on \textit{in situ} cleaned ZnO substrates. The structural analysis by LEED showed a growth in (111) direction and long-range order of the films. The comparison to films grown on MgO as well as the vertical depth profiling of the chemical structure by HAXPES confirmed a single-phase growth of the entire film. The uniform growth was achieved by varying the oxygen pressure during growth to avoid a reduced (FeO-like) interface and also the formation of Fe$_{2}$O$_{3}$ during film growth. Both kinds of films display in-plane easy magnetization with little if any in-plane magnetic anisotropy. The promising results obtained in this study will be followed by a detailed analysis of the relaxation mechanism that must happen in the first monolayers of the interface as well as spin transport measurements of such hybrid structures.

\begin{acknowledgments}
The authors gratefully acknowledge the technical support at BESSY. A. M\"uller also acknowledges financial support by the European Science Foundation (ESF), Thin Films for Novel Oxide Devices (THIOX). This work was funded by the BMBF (grant 05KS7WW3) and completed within the DFG Research Unit FOR 1162 (Fa 222/5-1 and Si 851/1-1).
\end{acknowledgments}
%

\end{document}